\documentstyle[preprint,tighten,aps,psfig]{revtex}

\begin{document}
\title{CHIRAL SYMMETRY RESTORATION IN HADRON SPECTRA
\footnote{Invited talk given at International School on Nuclear Physics
"Quarks in Hadrons and Nuclei",16th -24th September 2002, Erice/Sicily/Italy;
to appear in Progr. Part. Nucl. Phys., vol. 50}}
\author{ L. Ya. GLOZMAN }
\address{ Institute for Theoretical
Physics, University of Graz, Universit\"atsplatz 5, A-8010
Graz, Austria}
\maketitle

\begin{abstract}
{The evidence and the theoretical justification of  chiral
  symmetry restoration in high-lying hadrons is presented.}
\end{abstract} 

\section{Chiral symmetry restoration of the second kind}

It has recently been suggested that the parity doublet
structure seen in the spectrum of highly excited baryons
may be due to effective chiral symmetry restoration for
these states \cite{G1}. This phenomenon can be
understood in very general terms from  the validity of the operator
product expansion (OPE) in QCD at large space-like momenta and
the validity of the dispersion relation for the two-point
correlator, which connects the spacelike and timelike
regions (i.e. the validity of K\"allen-Lehmann representation)
\cite{CG1,CG2}.

Consider a two-point correlator $\Pi_{J_\alpha}$ of the current 
$J_\alpha(x)$ 
(that creates from 
the vacuum the hadrons with the  quantum numbers  $\alpha$) at large
spacelike momenta $Q^2$, where the language of quarks and
gluons is adequate and where the OPE is valid. The only effect that chiral
symmetry breaking can have on the correlator is through the nonzero
value of condensates associated with operators which are chirally
active ({\it i.e.} which transform nontrivially under chiral
transformations). To these belong $\langle  \bar q q \rangle$
and higher dimensional condensates that are not invariant under
axial transformation.  At large $Q^2$ only a small number of 
condensates need be retained to get an accurate description of the 
correlator. Contributions of these condensates are suppressed
by inverse powers of $Q^2$. At asymptotically
 high $Q^2$, the correlator is well described by a single term---the
 perturbative term.  The
 essential thing to note from this OPE analysis is that the perturbative
 contribution knows nothing about chiral symmetry breaking as it contains
 no chirally nontrivial condensates. In other words, though the chiral
 symmetry is broken in the vacuum and all chiral noninvariant condensates
 are not zero, their influence on the correlator at asymptotically
 high $Q^2$ vanishes. This is in contrast to the situation of low values of
 $Q^2$, where the role of chiral condensates is crucial.
 
 This shows that at large spacelike momenta  the
 correlation function becomes chirally symmetric.
 In other words, two correlators $\Pi_{J_1}(Q^2)$ and $\Pi_{J_2}(Q^2)$,
 where $ J_1 = UJ_2U^\dagger$, $U \in SU(2)_L \times SU(2)_R$,
 become essentially the same at large $Q^2$.
  The
 dispersion relation provides a connection between the
 spacelike and timelike domains. In particular, the large
 $Q^2$ correlator is completely dominated by the large $s$
 spectral density $\rho(s)$. (The spectral density  has the physical 
 interpretation 
 of being proportional to the
probability density that the current  when acting on the vacuum
creates a state of a mass of $\sqrt{s}$.) Hence the large $s$ spectral
density must be insensitive to the chiral symmetry breaking in
the vacuum. I.e. $\rho_1(s)$ and $\rho_2(s)$ must coincide at
asymptotically large $s$.
 This is in contrast to the low $s$ spectral functions
which are crucially dependent on the quark condensates in the
vacuum. This manifests a smooth chiral symmetry restoration
from the low-lying spectrum, where the chiral symmetry breaking
in the vacuum is crucial for physics, to the high-lying spectrum,
where chiral symmetry breaking becomes irrelevant and the spectrum
is chirally symmetric.

Microscopically this is because
the typical momenta of valence quarks should increase
higher in the spectrum and once it is high enough the
valence quarks decouple from the chiral condensates of
the QCD vacuum and the dynamical (quasiparticle or constituent)
mass of quarks drops off and the chiral symmetry gets restored
\cite{G1,G2}. This phenomenon does not mean that the spontaneous
breaking of chiral symmetry in the QCD vacuum  disappears,
but rather that the chiral asymmetry of the vacuum becomes
irrelevant  sufficiently high in the spectrum. 
The physics of the highly excited states is such as if there
were no chiral symmetry breaking in the vacuum.
One of the consequences is that the concept of constituent
quarks (which are directly related to the quark condensates
of the QCD vacuum), which is adequate low in the spectrum, becomes
irrelevant high in the spectrum.

How should one refer to such a phenomenon? Typically under
chiral symmetry restoration people understand that the chiral
properties of the vacuum are changed with temperature or/and density.
E.g. at  critical temperature the phase transition occurs
and the quark vacuum becomes trivial: all the quark condensates
of the vacuum vanish. On the contrary, in our case one
does not affect the QCD vacuum by exciting the hadrons. The
symmetry restoration is achieved via different mechanism. Namely,
by exciting the hadrons one {\it decouples} the valence degrees of
freedom from the QCD vacuum. So even if the quark condensates
of the vacuum (which break the chiral symmetry) are not zero, their
role gets smaller and smaller once we go up in the spectrum and
smoothly the chiral symmetry gets restored. 
 We can refer to such a phenomenon as chiral symmetry
restoration of the second kind.

All this is in a very good analogy with the similar
phenomenon in condensed matter physics. Consider a metal
which is in a superconducting phase. There is a
condensation of the Cooper pairs - which is analogous
to the condensation of right-left quark pairs in the QCD
vacuum - and as a consequence the low-lying excitations of the system are
the excitations of  quasiparticles - which are analogous
to the constituent quarks in the low-lying hadrons. Now
we want to study this superconductor by external probe,
e.g. by photons. If we probe the superconductor by the low-energy
photons $\hbar \omega \sim \Delta$, then the condensation
of the Cooper pairs and the quasiparticle structure of
the low-lying excitations are of crucial importance. One
clearly sees a gap $\Delta$ and a quasiparticle structure
in the response functions. However,
if one probes the same superconductor by the high-energy
photons $\hbar \omega \gg \Delta$, then the response of the
superconductor is the same as of the normal metal - the
high energy photons do not see quasiparticles and instead
they are absorbed by the bare electrons. This is because the
long-range phase coherent correlations in the superconductor
 become irrelevant in this case and the external probe sees
 a bare particle rather than a quasiparticle. Similar,
 in the QCD case in order to create a hadron of a large mass
 one has to probe the QCD vacuum by the high energy (frequency)
 external probe (current) and hence the physics (masses) of
 the highly excited hadrons should be insensitive to the
 condensation of the chiral pairs in the vacuum.
 
 \section{A simple pedagogical example}

It is instructive to consider a very simple quantum
mechanical example of symmetry restoration high in
the spectrum. Though there are conceptual differences
between the field theory with spontaneous symmetry
breaking and the one-particle quantum mechanics (where
only explicit symmetry breaking is possible), nevertheless
this simple example illustrates how this general phenomenon
comes about.

The example we consider is a two-dimensional harmonic
oscillator. We choose the harmonic oscillator only
for simplicity; the property that will be discussed below
is quite general one and can be seen in other systems.
The Hamiltonian of the system is invariant under 
$U(2) = SU(2)\times U(1)$ transformations. This
symmetry has profound consequences on the spectrum of the system.
 The energy levels of this  system are trivially found and
  are given by
\begin{equation}
E_{N, m} \, = \, ( N \, + \, 1 ); ~ m \, =
\, N, N-2, N-4, \, \cdots \, , -(N-2) , -N \; ,
\label{hoeigen}\end{equation}
where $N$ is the principle quantum number  and m is the
(two dimensional) angular momentum.  As a consequence of the symmetry, 
the levels are $(N+1)$-fold
degenerate.

Now suppose we add to the Hamiltonian a $SU(2)$ symmetry breaking
interaction (but which is still $U(1)$ invariant) of the form

\begin{equation}
V_{\rm SB} \, = \, A \, \theta (r - R),
\label{vsb}
 \end{equation}

\noindent
where  $A$ and $R$ are parameters and $\theta$ is the step
function.  Clearly, $V_{\rm SB}$ is not invariant under the
$SU(2)$ transformation. Thus the $SU(2)$ symmetry
is explicitly broken by this additional interaction, that acts
only within a circle of radius $R$.
As a result one would expect that the eigenenergies will not
reflect the degeneracy structure of seen in eq.~(\ref{hoeigen})
 if the
coefficients $R,A$ are sufficiently large.  Indeed,
 we have solved
numerically for the eigenstates for the case of $A=4$ and $R=1$
in  dimensionless units and one does not see a
multiplet structure in the low-lying spectrum as can be seen in
Fig.~1.

What is interesting for the present context is the high-lying spectrum.
  In Fig.~1 we have also plotted the energies between 70 and 74 for
   a few of the lower $m$'s.
   A multiplet structure is quite evident---to very good approximation
    the states of different $m$'s form degenerate multiplets and,
    although we have not shown this in the figure these multiplets
     extend in $m$ up to $m=N$.  
     
     How does this happen? The symmetry breaking
      interaction  plays a dominant role in the
       spectroscopy for small energies. Indeed, at small
       excitation energies
       the system is mostly located at distances where the symmetry
       breaking interaction acts and where it is dominant.
Hence  the low-lying spectrum to a very large extent is motivated  
by the symmetry breaking interaction.   However, at high 
excitation energies
the system mostly lives at large distances, where physics is dictated
by the unperturbed harmonic oscillator.  Hence at
 higher energies the spectroscopy
       reveals the $SU(2)$ symmetry of the two-dimensional harmonic
        oscillator.

\begin{figure}
\hspace*{-0.5cm}
$\begin{array}{cc}\psfig{file=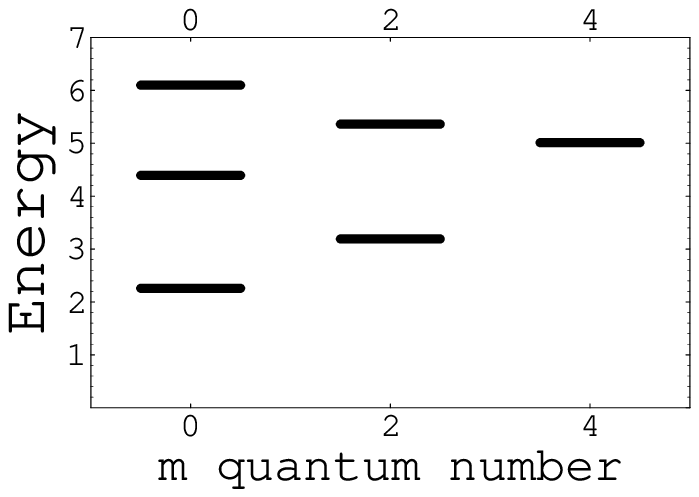,width=0.5\textwidth}
&\hspace*{-0.5cm}\psfig{file=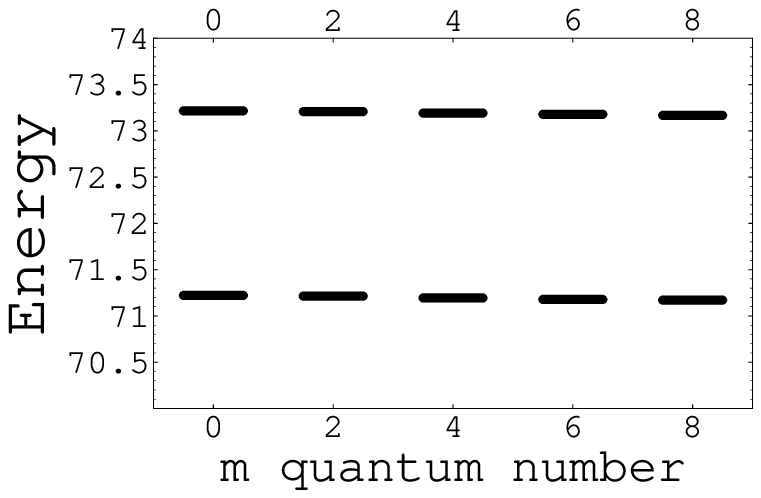,width=0.5\textwidth}\end{array}$
\caption{The low-lying (left panel) and highly-lying (right panel)
spectra of two-dimensional harmonic oscillator with the 
$SU(2)$-breaking term.}
\end{figure}

\section{Implications for baryon spectra}

If chiral symmetry restoration of the second kind
happens in the regime where the spectrum is still
quasidescrete (i.e. the successive excitations with
the given spin are well separated), then the symmetry
restoration imposes very strong constraints on the
spectrum, which will be discussed below.  A question 
then arises as to which
extent and where the hadron spectrum is still quasidescrete?
Clearly it is the case for the low-lying resonances. However,
the resonances with the given spin are still well separated
in the mass region $M \sim 2$ GeV and higher, which can be seen
from the nucleon spectrum, see Fig. 2, as well as from the meson
spectrum \cite{BUGG1,BUGG2}. In addition, if the linear-like
behaviour of both angular ($M^2 \sim J$) and radial ($M^2 \sim n$,
 where $n$ numerates radial excitations of the states
 with the given spin)
 Regge trajectories is valid up to large masses, then the spectrum 
 should be still quasidescrete in the region of validity of
 Regge phenomenology.

What are the implications of the chiral symmetry restoration
for a quasidiscrete spectrum? The equality of the spectral
functions $\rho_{1}(s)$ and $\rho_{2}(s)$ means that
 both masses $m_1$ and $m_2$ 
 as well as the amplitudes 
 $ \langle 0 | J_1 | n_1 \rangle$ and $\langle 0 | J_2 | n_2 \rangle $
 coincide for all the successive radial excitations.
In other words, the excited states must fill
in the irreducible representations of the parity-chiral
group \cite{CG1,CG2}. Thus the task is to find all possible 
representations of the $SU(2)_L \times SU(2)_R$ group that
are compatible with the definite parity of the physical state.
We emphasize parity because the irreducible representations
of the chiral group $(I_L,I_R)$ ( with $I_L$ and $I_R$ being 
isospins of the left and right subgroups) are not generally
 eigenstates of the parity operator, because under parity
 operation the left quarks transform into the right ones and
 vice versa.  However, a direct sum of two irreducible representations
 $(I_L,I_R) \oplus (I_R,I_L)$ does contain  eigenstates of the parity
 operator and hence the corresponding multiplet should be considered
 as a set of basis states for physical hadrons. What  crucially important
is that such a multiplet necessarily includes baryons with opposite
 parity.

Since the total isospin $I$ of a baryon can be obtained from
the left and right isospins $I_L$ and $I_R$ according to a standard angular
momentum addition rules and since there are no baryons with isospin
greater than 3/2, one immediately obtains the following allowed
parity-chiral multiplets $(1/2,0) \oplus (0,1/2)$, 
$(3/2,0) \oplus (0,3/2)$, $(1/2,1) \oplus (1,1/2)$. The first multiplet
corresponds to parity doublets of any spin in the nucleon spectrum.
The second one describes the parity doublets of any spin in
the delta spectrum. However, the latter multiplet 
combines one parity doublet in the nucleon spectrum with the
parity doublet in the delta spectrum with the same spin.

Summarizing, the phenomenological consequence  of the chiral symmetry
restoration of the second kind
high in $N$ and $\Delta$ spectra is that the baryon states
will fill out  the irreducible
representations of the parity-chiral group.
If $(1/2,0) \oplus (0,1/2)$ and $(3/2,0) \oplus (0,3/2)$
multiplets were realized in nature, then the spectra of highly excited
nucleons and deltas would consist of parity doublets. However,
 the parity doublet with  given spin in
the nucleon spectrum {\it a-priori} would not be degenerate with the
 doublet with the same spin in the delta spectrum;
these doublets would belong to different
representations , {\it i.e.} to distinct
multiplets and their energies
are not related.   On the other hand,
if $(1/2,1) \oplus (1,1/2)$ were realized, then the high-lying
 states in $N$ and $\Delta$ spectrum
would have a $N$ parity doublet and a $\Delta$
parity doublet with the same spin and which are degenerate in mass.
In either of cases the high-lying spectrum  must systematically
consist of parity doublets. We stress that this classification is the 
most general one
and does not rely on any model assumption about the
structure of baryons. 
 What  is interesting  is that the same classification
can be trivially obtained if one assumes that the
chiral properties of baryons in the chirally restored regime
are determined by three {\it massless} valence quarks.

Now we have to analyze an empirical spectrum of baryons
in order to see whether there are  signs of
chiral symmetry restoration. 
What is immediately evident from the empirical
low-lying spectrum is
that  positive and negative
parity states with the same spin are not nearly degenerate.
Even more, there is no one-to-one mapping of positive and
negative parity states of the same spin with masses below 1.7 GeV.
This means that one cannot
describe the low-lying spectrum as consisting of sets of chiral
partners. It is not so surprising since  the low-lying spectrum is
mostly driven by the chiral symmetry breaking effects.
The absence of   parity doublets low in the spectrum
 is one of the most direct pieces of evidence that chiral symmetry in QCD is
spontaneously  broken (and very strongly).

\begin{figure}
\hspace*{-0.5cm}
\centerline{
\psfig{file=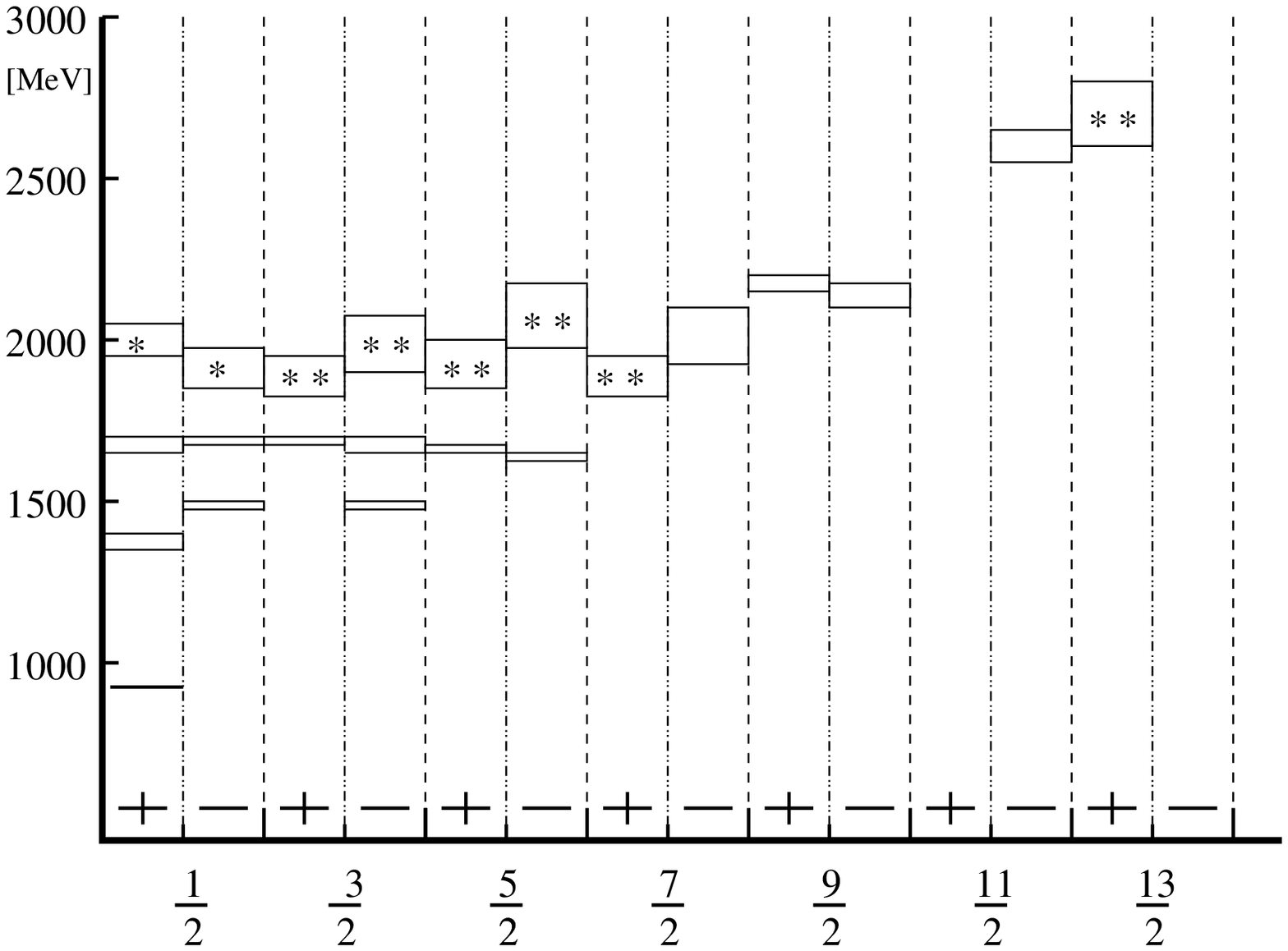,angle=-90,width=0.6\textwidth}
}
\caption{Excitation spectrum of the nucleon. The real part
of the pole position is shown. Boxes represent experimental
uncertainties. Those resonances which are not yet established
are marked by two or one stars according to the PDG classification.
The one-star resonances with $J=1/2$ around 2 GeV are given
according to the recent Bonn (SAPHIR) results. }
\end{figure}

Starting at the mass $M \sim 1.7$ GeV one observes three
almost perfect parity doublets with spins $J=1/2,3/2,5/2$.
It is important that all these states are well established
and have **** or *** status according to PDG classification.
The next excitations with the same quantum numbers are not yet
established though existing crude data support parity doubling
phenomenon. The lowest excitations with spin 9/2 are also
well established
states and they represent another good example of parity doubling.
There are well established states with $J=7/2$ and $J=11/2$ where
the chiral partners have not yet been identified. According
to a chiral symmetry restoration scenario they must exist.
So it is a very interesting and important experimental task
to find such states as well as to clear up a situation for
$J=1/2,3/2,5/2$ resonances around 2 GeV.

In the delta spectrum there are also obvious parity doublets
starting from the mass region 1.9 - 2 GeV. Again, there are
well established states (like $J^P=7/2^+$ at 1950 MeV) where
the chiral partner has not yet been identified. So similar to
the nucleon spectrum, the upper part of the delta spectrum
must be  experimentally explored. 

\section{Implications for meson spectra}

Recent data on  highly excited
mesons also suggest an evidence for chiral symmetry
restoration in hadron spectra   \cite{G2}. Consider, as an
example, the pseudoscalar and scalar mesons $\pi, f_0,a_0,\eta$
within the two-flavor QCD. The corresponding currents (interpolating
fields) $J_\pi(x)$, $J_{f_0}(x)$,$J_{a_0}(x)$ and $J_\eta(x)$
belong to the $(1/2,1/2) \oplus (1/2,1/2)$ irreducible representation
of the 
$U(2)_L\times U(2)_R = SU(2)_L\times SU(2)_R \times U(1)_V\times U(1)_A$
group. Specifically the pairs $(J_\pi(x)$, $J_{f_0}(x))$ and 
$(J_{a_0}(x),J_\eta(x))$ form the basis of the  $(1/2,1/2)$
representation of the chiral group $SU(2)_L\times SU(2)_R$.
If the vacuum were invariant with respect to $U(2)_L\times U(2)_R $
transformations, then all four mesons, $\pi,f_0,a_0$ and $\bar \eta$
would be degenerate (as well as all their excited states). Once
the $U(1)_A$ symmetry is broken explicitly through
the axial anomaly, but the chiral $SU(2)_L\times SU(2)_R $
symmetry is still
intact in the vacuum, then the spectrum would consist of
degenerate $(\pi, f_0)$ and $(a_0, \bar \eta)$ pairs.\footnote
{$\bar \eta$ represents the two-flavor singlet state; its
mass for the lowest mesons can be extracted from the
physical $\eta$ and $\eta'$ states.}
 If
in addition the chiral  $SU(2)_L\times SU(2)_R $ symmetry is
spontaneously broken
in the vacuum, the degeneracy is also lifted in  the pairs
above and the pion becomes a (pseudo)Goldstone boson. Indeed,
the masses of the lowest mesons  are 
 
 $$ m_\pi \simeq 140 MeV, ~m_{f_0} \simeq 400 - 1200 MeV,~
m_{a_0} \simeq 985 MeV ,~ m_{\bar \eta} \simeq 782 MeV. $$

This immediately tells that both $SU(2)_L\times SU(2)_R $ and
$U(1)_V \times U(1)_A$ are broken in the QCD vacuum
 to $SU(2)_I$ and $U(1)_V$, respectively.
 
Systematic data on highly excited mesons are still missing
in the PDG tables. We will use  recent
results of the partial wave analysis of mesonic resonances
from 1.8 GeV to 2.4 GeV
obtained in $p \bar p$ annihilation at LEAR, see Table below.
We note that the $f_0$ state at $2102 \pm 13$  MeV is {\it not} 
 considered by the authors as  a $q\bar q$ state (but rather as a
candidate for glueball) because of its very unusual decay
properties and very large mixing angle. This is in contrast to all
other $f_0$ mesons in this region, for which  the mixing angles
are small. Therefore these mesons are regarded  as
predominantly $u,d = n$ states.
 Hence, in the following we will
exclude the $f_0$ state at $2102 \pm 13$  from our analysis which
applies only to $n\bar n$ states. 

\begin{center}
\begin{tabular}{|llllll|} \hline
Meson & ~I~ & $~J^P~$ & Mass (MeV) & Width (MeV) & Reference\\ \hline
$f_0$ & ~0~ & $~0^+~$  & $1770 \pm 12$ &  $220 \pm 40$ & \cite{BUGG0}\\
$f_0$ & ~0~ & $~0^+~$  & $2040 \pm 38 $ &  $405 \pm 40$ & \cite{BUGG1} \\
$f_0$ & ~0~ & $~0^+~$  & $2102 \pm 13$  &  $211 \pm 29$ & \cite{BUGG1} \\
$f_0$ & ~0~ & $~0^+~$  & $2337 \pm 14$  &  $217 \pm 33$ & \cite{BUGG1} \\
$\eta$ & ~0~ & $~0^-~$  & $2010^{+35}_{-60}$   &  $270 \pm 60$ & \cite{BUGG1}\\
$\eta$ & ~0~ & $~0^-~$  & $2285 \pm 20$   &  $325 \pm 30$ & \cite{BUGG1} \\
$\pi$ & ~1~ & $~0^-~$  & $1801 \pm 13$   &  $210 \pm 15$ & \cite{PDG} \\
$\pi$ & ~1~ & $~0^-~$  & $2070 \pm 35$   &  $310^{+100}_{-50}$ & \cite{BUGG2}\\
$\pi$ & ~1~ & $~0^-~$  & $2360 \pm 25$   &  $300^{+100}_{-50}$ & \cite{BUGG2}\\
$a_0$ & ~1~ & $~0^+~$  & $2025 \pm ?$   &  $320 \pm ?$ & \cite{BUGG2}\\
 \hline
\end{tabular}
\end{center}

 The prominent feature of the data is an approximate
degeneracy of the three highest states in the pion spectrum with
the three highest states in the $f_0$ spectrum:
 
\begin{equation}
\pi(1801 \pm 13) - f_0(1770 \pm 12),
\end{equation}
 
\begin{equation}
\pi(2070 \pm 35) - f_0(2040 \pm 38 ),
\end{equation}
 
\begin{equation}
\pi(2360 \pm 25) - f_0(2337 \pm 14).
\end{equation}

 This can be considered as  a manifestation
 of chiral symmetry restoration high in the spectra. The approximate
 degeneracy of these physical states indicates that the
 chiral $SU(2)_L \times SU(2)_R$ transformation properties
 of the corresponding currents  are not violated
 by the vacuum. This means that the chiral symmetry breaking of 
 the vacuum becomes
 irrelevant for the high-lying states and the physical states
 above form approximately the  chiral pairs
 in the $(1/2,1/2)$ representation of the chiral group.
The physics of the high-lying hadrons is such as if there
were no spontaneous chiral symmetry breaking.

 A similar behaviour is observed from a comparison of the $a_0$ and
 $\eta$ masses high in the spectra:
 
 \begin{equation}
 a_0(2025 \pm ?) - \eta(2010^{+35}_{-60}).
\end{equation}

Upon examining the experimental data more carefully one notices
not only a degeneracy in the chiral pairs, but also an approximate
degeneracy
in $ U(1)_A $ pairs $(\pi, a_0)$ and $(f_0, \eta)$
(in those cases where the states are established).
If so, one can preliminary conclude that not only the
chiral $SU(2)_L\times SU(2)_R $ symmetry is restored, 
but the whole $U(2)_L\times U(2)_R $ symmetry of
the QCD Lagrangian. Then the approximate $(1/2,1/2) \oplus (1/2,1/2)$
multiplets of this group are given by:
 
\begin{equation}
 \pi(1801 \pm 13) - f_0(1770 \pm 12) - a_0 (?) -
 \eta(?);
\label{quartet1}
\end{equation}

\begin{equation}
 \pi(2070 \pm 35) - f_0(2040 \pm 40) - a_0 (2025 \pm?) -
 \eta(2010^{+35}_{-60});
 \end{equation}
 
 \begin{equation}
 \pi(2360 \pm 25) - f_0(2337 \pm 14) - a_0(?) -
 \eta(2285 \pm 20).
\label{quartet3}
\end{equation}
 
\noindent
 
 This
preliminary conclusion would be strongly supported by a discovery
of the missing $a_0$ meson in the mass region around 2.3 GeV
as well as by the missing $a_0$ and $\eta$ mesons in the 1.8 GeV region.
That these missing mesons should indeed exist is also
supported by the hypothesis of the linear radial Regge
trajectories for highly excited states \cite{BUGG1,BUGG2}.
We have to stress, that the $U(1)_A $ restoration
high in the spectra does not mean that the axial anomaly
of QCD vanishes, but rather that the specific gluodynamics
(e.g. instantons) that are related to the anomaly
become unimportant there. It should also be emphasized
that the only restoration of  $U(1)_V \times U(1)_A $ symmetry
(without the $SU(2)_L\times SU(2)_R $) is impossible. This was discussed
in ref. \cite{CG1}. The reason is that even if the effects
of the explicit $ U(1)_A $ symmetry breaking via the axial anomaly
vanish, the $U(1)_V \times U(1)_A $ would  still be spontaneously
broken once the $SU(2)_L\times SU(2)_R $ were spontaneously broken.
This is because the same quark condensates in the QCD vacuum
that break $SU(2)_L\times SU(2)_R $ do also break $U(1)_V \times U(1)_A $.\\

\section{Implications  for 
modeling of hadrons}

It is quite natural to assume that the physics of the
highly excited hadrons is due to confinement in QCD.
If so it follows that the confining gluodynamics is
still important in the regime where chiral symmetry
breaking in the vacuum has become irrelevant. Then
it follows that the mechanisms of confinement and
chiral symmetry breaking in QCD are not the same.

The phenomenon 
of chiral symmetry restoration high in the spectra
rules out the potential description of high-lying hadrons in the
spirit of the constituent quark model. Clearly, the chiral symmetry 
restoration by itself implies that constituent
quarks as effective degree of freedom (whose mass is directly
 related to spontaneous
chiral symmetry breaking in the vacuum) become inadequate
high in the spectrum, though it is a fruitful concept for the
low-lying hadrons. That the potential description is incompatible
with the parity doubling is also seen from the following.

Consider, for instance, mesons.
Within the potential description of mesons the parity of the state is
unambiguously prescribed by the relative orbital angular
momentum $L$ of  quarks. For example,
 all the  states  on the radial pion Regge trajectory are
$^1S_0$  $q \bar q$ states, while the members of the $f_0$ trajectory
are the $^3P_0$ states. Clearly, such a picture cannot
explain the {\it systematical} parity doubling as it would require
that the stronger centrifugal repulsion in the case of $^3P_0$
mesons (as compared to the $^1S_0$ ones) as well as the strong and attractive
 spin-spin force in the case of $^1S_0$ states (as compared to
 the weak  spin-spin force in the $^3P_0$ channel)
 must systematically lead to an approximate degeneracy for all
 radial states. This is very improbable. Similar conclusions
 can be easily obtained for baryons \cite{G1}. More generally,
 the chiral symmetry restoration of the second kind is in
 contradiction with all models where chiral symmetry breaking
 is induced by  confinement.

  The potential
picture also implies  strong spin-orbit interactions
between quarks while the spin-orbit splittings are absent
or very small for excited mesons and baryons in the $u,d$ sector.
The strong spin-orbit interactions inevitably follow from the
Thomas precession (once the confinement is described through a scalar
confining potential)\footnote{
Note also that a scalar potential explicitly breaks
the chiral symmetry
in contradiction to the requirement that the chiral symmetry must
be restored high in the spectra.},
and this very strong spin-orbit force must be practically
exactly compensated by  other strong spin-orbit force from  e.g. the
one-gluon-exchange interaction in this picture.
In principle such a cancellation could be provided by   tuning
the parameters for some specific (sub)families of mesons. However,
in this case the spin-orbit forces become very strong for  other
(sub)families. This is a famous spin-orbit problem of constituent
quark model.

This picture should be contrasted with the string description
of highly excited hadrons \cite{G3}. Within the latter
one these hadrons are the relativistic strings (with
the color-electric
field in the string) with practically massless bare quarks
at the ends;  these massless quarks are combined into
parity-chiral multiplets.
The string picture is compatible with the chiral symmetry
restoration because there always exists a solution for the
right-handed and left-handed quarks at the end of the string
with exactly the same energy and total angular momentum.
Since the nonperturbative field in the string is pure electric
and the electric field is "flavor-blind",
the string dynamics itself is not sensitive to the specific
flavor of a light quark once the chiral limit is taken.
 This picture
explains the empirical
parity-doubling because for every intrinsic quantum state
of the string there necessarily appears parity doubling
of the states with the same total angular momentum of hadron.
Hence the string picture is compatible not only  with the
$SU(2)_L \times SU(2)_R$ restoration, but more generally
with the $U(2)_L \times U(2)_R$ one. In addition, there is
no spin-orbit force at all once the chiral symmetry is restored.
This is because the helicity operator does not commute with
the spin-orbit operator and a motion of a quark with
the fixed helicity is not affected by the spin-orbit force.

\end{document}